\documentstyle[12pt,aasms]{article}
\newcommand{\bqa}{\begin{eqnarray*}}
\newcommand{\eqa}{\end{eqnarray*}}
\newcommand{\bqan}{\begin{eqnarray}}
\newcommand{\eqan}{\end{eqnarray}}

\newcommand{\vV}{{\bf V}}
\def\etal{{\it et~al.\ }}

\tightenlines

\begin{document}
 
\title{Linear Regression for Astronomical Data with Measurement
Errors and Intrinsic Scatter}
 
\author{Michael G. Akritas}
\affil{Department of Statistics,
The Pennsylvania State University, State College PA 16802 \\
mga@stat.psu.edu}
\and
\author{Matthew A. Bershady\altaffilmark{1}}
\affil{Department of Astronomy \& Astrophysics,
The Pennsylvania State University, State College PA 16802\\
mab@astro.psu.edu}

\altaffiltext{1}{Hubble Fellow}

\vskip 2in 

\centerline{To appear in the {\it Astrophysical Journal}, October 10,
1996 issue, Vol. 470}

\vskip 2in 

\begin{abstract}
 
Two new methods are proposed for linear regression analysis for data
with measurement errors. Both methods are designed to accommodate
intrinsic scatter in addition to measurement errors. The first method
is a direct extension of the ordinary least squares (OLS) estimator to
allow for measurement errors.  It is quite general in that a) it
allows for measurement errors on both variables, b) it allows the
measurement errors for the two variables to be dependent, c) it allows
the magnitudes of the measurement errors to depend on the
measurements, and d) other `symmetric' lines such as the bisector and
the orthogonal regression can be constructed.  We refer to this method
as BCES estimators (for Bivariate Correlated Errors and intrinsic
Scatter). The second method is a weighted least squares (WLS)
estimator, which applies only in the case where the `independent'
variable is measured without error and the magnitudes of the
measurement errors on the 'dependent' variable are independent from
the measurements. 

Several applications are made to extragalactic astronomy: The BCES
method, when applied to data describing the color-luminosity relations
for field galaxies, yields significantly different slopes than OLS and
other estimators used in the literature. Simulations with artificial
data sets are used to evaluate the small sample performance of the
estimators.  Unsurprisingly, the least-biased results are obtained
when color is treated as the dependent variable. The Tully-Fisher
relation is another example where the BCES method should be used
because errors in luminosity and velocity are correlated due to
inclination corrections.  We also find, via simulations, that the WLS
method is by far the best method for the Tolman surface-brightness
test, producing the smallest variance in slope by an order of
magnitude. Moreover, with WLS it is not necessary to ``reduce''
galaxies to a fiducial surface-brightness, since this model
incorporates intrinsic scatter.
 
\end{abstract}

 
\section{Introduction}
 
Linear regression analysis is used extensively in everyday
astronomical research. The distinguishing feature of many astronomical
data sets is the presence of intrinsic scatter in addition to
heteroscedastic measurement errors (i.e. the size of the error can
vary from observation to observation). A few notable examples in
extra-galactic astronomy include relations between X-ray temperatures
and velocity dispersions for galaxy clusters, the color-luminosity
relations for field galaxies, the Tully-Fisher relation (and other
``Fundamental Plane'' relations), and the Tolman test. In this paper
we consider the latter three examples. Neglect of measurement errors
and intrinsic can bias the derived slopes of these relations, thus
potentially leading to incorrect astrophysical deductions. Willick
(1991) has performed a detailed study of the effects of intrinsic
scatter and measurement error on the estimated slope of the
Tully-Fisher relation. An additional issue pertains to biases
associated with the correlation of measurement errors between
observables. This occurs, for example, in color-luminosity relations
and the Tully-Fisher relation. There are some cases where the
measurement error is negligible in one variable.  For example, in the
case of the Tolman surface-brightness test, the error in redshift is
usually negligible compared to the error and intrinsic scatter in
surface-brightness.

Regression model found in statistics text books rarely accommodate
heteroscedastic measurement errors in more than one variable, and the
recent paper Isobe \etal (1991) (IFAB hereafter) deals exclusively
with data that have no measurement errors. Indeed, accommodating
heteroscedastic measurement errors and intrinsic scatter is mentioned
in Feigelson \& Babu (1992) as one of the outstanding problems in
linear regression.

The only currently available regression methods that deal with
heteroscedastic measurement errors are based on the assumption that
the true variables (in the absence of measurement error) have no
intrinsic scatter. That is, the true points are assumed to lie
exactly on a straight line, which implies they have correlation one.
Software packages which perform regressions under this assumption are
mentioned in Feigelson \& Babu (1992), including ORDPACK (Boggs \etal
1990), which also does nonlinear regression. A more accessible
reference is Press {\it et al.} (1988). This assumption, however, is
violated in many astronomical data sets.

In this paper we address the important problem of fitting regression
models with data having heteroscedastic measurement errors of known
standard deviation, and entirely unknown intrinsic scatter.  We define
a {\it statistical model} for data with astronomical (heteroscedastic)
measurement errors which allows the possibility of correlated errors
between both variables of interest, and the possibility that the size
of the measurement error depends on the observation.  This model
should prove useful for addressing other problems with such data,
including intrinsic variance function estimation, goodness-of-fit,
comparing $k$ multivariate samples etc). Very important is the
distinction we draw between the case where the size of the measurement
error (standard deviation in statistical parlance) depends on the
measurement and the case where it does not. Both cases are equally
common in astronomical data sets (e.g. background-limited versus
source-limited observations). However, procedures that weigh the
measurements according to the variance of the measurement error can
give biased results if this variance depends on the observation, as we
discuss in the next section.

We describe here two different regression methods. Both of our
methods pertain only to linear (as opposed to nonlinear) regression
and are based on transparent ideas that make them very intuitive.
The first method is a direct generalization of the OLS estimator
which applies  quite generally.  The second method is a
weighted least squares (WLS) estimator which applies when only the
`response' variable is subject to measurement error and the size of
the measurement error does not depend on the observation. 

We only consider simple linear regression here (i.e.\ only one
`explanatory' variable); extensions of this method to multiple
regressions will appear in a sequel paper. The paper is organized as
follows. In the next section we introduce the basic idea of our
method. The statistical model for data with measurement errors is
presented in subsection 2.1. In subsection 2.1 we consider the general 
case where both the response and the explanatory variable are subject 
to potentially correlated measurement errors, and the magnitude of 
these errors may
depend on the measurements. We use the acronym BCES($X_2|X_1$) to
denote the present generalization of the OLS($X_2|X_1$), which
minimizes the residuals in $X_2$ (i.e. $X_2$ is the dependent
variable). In subsection 2.2 we consider the case where only the
response variable is subject to measurement error whose magnitude does
not depend on the measurement, and we introduce a competing procedure
based on WLS. In Section 3 we study other versions of the first
method, namely the BCES-bisector and BCES-orthogonal regression; these
regression lines are defined in terms of BCES($X_2|X_1$) and
BCES($X_1|X_2$). In Section 4 we apply these methods to an
astronomical data set and use simulations as a methodological tool to
investigate the small-sample performance of the four BCES estimators
(for color-luminosity relations) and the WLS estimator (for the
Tolman test). Relevance to the Tully-Fisher relation is discussed. We
consider more general application of BCES in Section 5. The
mathematical derivations are given in the Appendix.
 
\newpage

\section{Simple Regression}

As explained in the introduction, observed astrophysical correlations
exhibit are expected to exhibit scatter due to both intrinsic and
measurement error. These two sources of scatter must be recognized and
separately treated to avoid biases. However, different kinds of
measurement errors have different effects -- some being benign. There
are two basic distinctions: a) whether or not the measurement error is
in the independent variable, and b) whether or not the magnitude of
the measurement error depends on the measurement.

It is well documented in the statistical literature that the OLS slope
is biased if there is measurement error in the independent variable.
Measurement errors only in the dependent variable are less critical:
Provided the measurement error is independent from the observation, 
OLS is valid but not efficient. The technical term "not efficient" 
means that there exists another estimator which has smaller scatter
(and therefore results in narrower confidence intervals). In particular,
if there is no measurement error in the independent variable and the 
measurement error in the dependent variable is independent from the 
measurement, the WLS estimator in subsection 2.3
is more efficient than the OLS estimator. On the other hand, if the 
size of the measurement error depends on the measurement, neither the 
OLS estimator nor the WLS estimator of subsection 2.3 are valid, even
if the independent variable is measured without error, and even if
there is no intrinsic scatter.
The technical term "not valid" means that the estimator is biased even for
large samples; as a consequence, the larger the sample size, the
less likely it will be for a confidence interval based on
it to contain the true value. The bias can get worse if there is
measurement errors in both variables, and more so if the measurement
errors are correlated.

In subsection 2.2 we present an extension of the OLS slope which is
valid in all cases. This is also the "conservative" estimator that
should be used when one is unclear about the prevailing conditions.
When the independent variable is measured without error and one is
certain that the magnitude of the measurement error does not depend on
the observation, the WLS estimator of subsection 2.3 is more efficient
and therefore recommended. First, however, we need to set the notation
and formulate a statistical model for data with heteroscedastic
measurement errors. This is done in subsection 2.1. Inadvertently,
these sections are technical.

\newpage

\subsection{A Statistical Model}

Let the variables of interest be denoted by $(X_{1i},X_{2i})$ and the
observed data be denoted by 
\bqan \label{data.simp.reg}
(Y_{1i},Y_{2i}, \vV_i), ~i=1,\ldots n, 
\eqan 
where for each $i$,
$\vV_i$ is a symmetric $2\times 2$ matrix with elements denoted by
$V_{11,i}, ~V_{22,i}$, and $V_{12,i}$, for the two
diagonal and the common off diagonal elements, respectively. The
observed data are related to the unobserved variables of interest by
\bqan \label{meas.error.mod} 
Y_{1i}=X_{1i} + \epsilon_{1i}, ~\mbox{and} ~Y_{2i}=X_{2i}+\epsilon_{2i}, 
\eqan 
where the errors
$(\epsilon_{1i},\epsilon_{2i})$ have a joint bivariate distribution
with zero mean and covariance matrix $\vV_i$, for all $i$. In this
model we allow $\vV_i$ to depend on $(Y_{1i},Y_{2i})$ and thus,
implicitly on $(X_{1i},X_{2i})$. Thus we do not 
require that $(\epsilon_{1i},\epsilon_{2i})$ be independent from 
$(X_{1i},X_{2i})$. However, we assume that $\vV_i$ is the only 
aspect of the distribution of $(\epsilon_{1i},\epsilon_{2i})$ that depends
on $(Y_{1i},Y_{2i})$. In other words, we assume that, given $\vV_i$,
$(\epsilon_{1i},\epsilon_{2i})$ is independent from $(X_{1i},X_{2i})$.

\medskip

The intuitive meaning of the technical assumption that "given $\vV_i$,
$(\epsilon_{1i},\epsilon_{2i})$ is independent from $(X_{1i},X_{2i})$"
is that $\epsilon_{1i}$, for example, is equally likely to be positive
or negative for any value of $X_{1i}$, and the size of its
absolute value is governed (in addition to the type of the 
measurement error distribution) by the magnitude of $V_{11,i}$ which
is given. All astronomical data sets that we are aware of comply to
this assumption.

In most cases, the measurement errors for the two variables are
independent (so $V_{12,i} =0$ for all $i$), and the observed data is
of the form \footnote{Very often, astronomical data sets will not give
explicitly the magnitude of the uncertainty of the errors (i.e.
$V_{11,i}, ~V_{22,i}$).  Instead the uncertainty is reported in the
form of $(1-\alpha)100$\% (e.g. 95\%) confidence intervals $Y_{1i}\pm
c_{1i}, ~Y_{2i}\pm c_{2i}$.  In this case the $V$'s can be recovered
from the relation $c_{ki} = z_{\alpha /2} \sqrt{V_{kk,i}}$, for
$k=1,2$, where $z_{\alpha /2}$ is the $(1-\alpha /2)100$-th percentile
of the standard normal distribution.}
$$(Y_{1i},Y_{2i},V_{11,i},V_{22,i}),$$ with $V_{kk,i}$ denoting the
variance of $\epsilon_{ki}, ~k=1,2$.
 
\medskip
 
It is assumed that the variables of interest follow the usual simple
regression model \bqan \label{simp.reg.mod} X_{2i} = \alpha_1 +
\beta_1 X_{1i} + e_i, \eqan where $e_i$ is assumed to have zero mean
and finite variance.  The terminology "intrinsic scatter" (or
"intrinsic dispersion") is commonly used to indicate the variance or
standard deviation of $e_i$.  We want to estimate the regression
coefficients $\alpha_1, ~\beta_1$ and also estimate the uncertainties
of these estimators using the data in (\ref{data.simp.reg}).
 
\subsection{The BCES($X_2|X_1$) Estimator}
 
The estimator BCES($X_2|X_1$) (see the Abstract and Introduction for
explanation of the acronym) to be proposed here is an extension of the
OLS($X_2|X_1$) estimator which is valid for all data sets that fit the
measurement error model specified in subsection 2.1.  This estimator
is based on the fact that the parameters of (\ref{simp.reg.mod}) are
related to the moments of the bivariate distribution of
$(X_{1i},X_{2i})$. In particular, \bqan \label{param.expr} \beta_1 =
\frac{C(X_{1i},X_{2i})}{V(X_{1i})}, ~\mbox{and} ~\alpha_1 = E(X_{2i})
- \beta_1 E(X_{1i}), \eqan where $C(X_{1i},X_{2i})$ denotes the
covariance of $X_{1i}$ and $X_{2i}$, $V(X_{1i})$ denotes the variance
of $X_{1i}$ and $E$ denotes expected value.  In the case of no
measurement errors, the OLS estimators are simply moment estimators,
so they are obtained by replacing the population moments in
(\ref{param.expr}) by sample moments. The proposed estimators
generalize the OLS estimators by replacing the population moments in
(\ref{param.expr}) by moment estimators obtained from the observed
data (\ref{data.simp.reg}).  These moment estimators are based on the
following results: For $k=1$ or 2 we have \bqan
\label{mean.rel} E(Y_{ki}) &=& E(X_{ki}) \\ \label{var.rel} V(Y_{ki})
&=& V(X_{ki}) + E(V_{kk,i}) \\ \label{cov.rel} C(Y_{1i},Y_{2i})
&=& C(X_{1i},X_{2i}) + E(V_{12,i}).  
\eqan
The proof of these results is given in the Appendix.
 
Using relations (\ref{mean.rel}), (\ref{var.rel}), (\ref{cov.rel}) and
relation (\ref{param.expr}) we can express the regression parameters
$\alpha_1 , ~\beta_1$ in terms of the population moments of the
observed data. Thus, \bqan
\label{param.expr.2} \beta_1 = \frac{C(Y_{1i},Y_{2i}) -
E(V_{12,i})}{V(Y_{1i}) - E(V_{11,i})}, ~\mbox{and} ~\alpha_1
= E(Y_{2i}) -\beta E(Y_{1i}).  
\eqan 
This relation suggests the
following extension of the OLS estimator to data with measurement
errors, 
\bqan \label{hat.beta} \hat \beta_1 &=&
\frac{\sum_{i=1}^n(Y_{1i}-\bar{Y_1})(Y_{2i}-\bar{Y_2})
-\sum_{i=1}^nV_{12,i}}{\sum_{i=1}^n(Y_{1i}-\bar{Y_1})^2 -
\sum_{i=1}^nV_{11,i}} \\ \label{hat.alpha} \hat \alpha_1 &=&
\bar{Y_2} -\hat \beta_1 \bar{Y_1}.  
\eqan
These are the BCES($X_2|X_1$) estimators of the slope and intercept
which generalize the OLS($X_2|X_1$) estimators. The way the 
BCES($X_2|X_1$) slope estimator adjusts for the presence of measurement
errors is quite obvious from (\ref{hat.beta}). Namely, if the errors are
correlated, the observed covariance between $Y_1$ and $Y_2$ is biased.
Therefore, the numerator of $\hat \beta_1$ consists of a ``debiased''
sample covariance. Similarly, the observed variance of $Y_1$ is biased 
necessitating the ``debiased'' sample variance seen in the denominator 
of $\hat \beta_1$.

It will be shown in the Appendix that these estimators all have, 
asymptotically, a zero mean normal distribution. To give expressions for
their variances, we need the following notation. Let
\bqan \label{xi_1}
\xi_{1i} &=& \frac{(Y_{1i}-E(Y_{1i}))(Y_{2i}-\beta_1 Y_{1i}-\alpha_1) +
\beta_1V_{11,i}-V_{12,i}}{V(Y_{1i}) - E(V_{11,i})}\\
\label{zeta_1}
\zeta_{1i} &=&Y_{2i} -\beta_1 Y_{1i} -E(Y_{1i})\xi_{1i},
\eqan
and let $\hat \xi_{1i}$, $\hat \zeta_{1i}$ be obtained by substituting 
the unknown quantities in $\xi_{i1}$, $\zeta_{1i}$ by their obvious 
estimators (i.e.\
substitute sample means in place of population means, sample variances
in place of population variances, and $\hat \beta_1 , ~\hat \alpha_1$ in
place of $\beta_1 , \alpha_1$). Finally, let $\bar{\hat\xi_1}$
($\bar{\hat\zeta_1}$) denote the arithmetic average of the 
$\hat \xi_{1i}$ ($\hat \zeta_{1i}$) and set
\bqan \label{sigma.beta.hat}
\hat \sigma^2_{\beta_1} &=& n^{-1}\sum_{i=1}^n(\hat\xi_{1i}
-\bar{\hat\xi_1})^2
\\
\label{sigma.alpha.hat}
\hat \sigma^2_{\alpha_1} &=&
n^{-1}\sum_{i=1}^n(\hat\zeta_{1i}-\bar{\hat\zeta_1})^2. 
\eqan 
Then the variance of $\hat\beta_1$ is estimated by
$\widehat V (\hat\beta_1) = n^{-1}\hat\sigma^2_{\beta_1}$; thus, the
asymptotic normality shown in the Appendix 
implies that a $(1-\alpha)100$\% confidence interval for $\beta_1$ is
\bqan \label{conf.inter} 
\hat \beta_1 \pm z_{\alpha/2}\hat\sigma_{\beta_1} n^{-1/2}.  
\eqan 
Similarly  the variance of $\hat\alpha_1$ is estimated by
$\widehat V (\hat\alpha_1) = n^{-1}\hat\sigma^2_{\alpha_1}$ with
a similar confidence interval implied from the asymptotic normality.

Finally let $\hat\sigma_{\beta_1,\alpha_1}$ be the sample covariance
obtained from $(\hat\xi_{1i},\hat\zeta_{1i})$. Then
the covariance between $\hat\beta_1$ and $\hat\alpha_1$ is estimated by
$$
\widehat{Cov}(\hat\beta_1,\hat\alpha_1) =
n^{-1}\hat\sigma_{\beta_1,\alpha_1}.
$$
This estimated covariance function can be used for constructing a 
simultaneous confidence ellipsoid for $\hat\beta_1$ and $\hat\alpha_1$.
See for example Johnson \& Wichern (1988).
 
\subsection{Only the Response Variable with Measurement Error}
 
In this subsection we describe a WLS estimator for the case that
$X_{1i}$ is observed without error. This estimator requires the 
additional assumption that the measurement error in $X_{2i}$ is 
independent of $X_{2i}$. The main idea behind this estimator is
that, in the presence of nonnegligible intrinsic scatter, the optimal 
weight for each observation is made up both from the variance of the
corresponding measurement error and the intrinsic scatter.
Thus, to determine the optimal weight we first need to estimate the 
intrinsic scatter. We proceed with a formal description of this method.
 
In the case that $V_{11,i} = 0$ for all $i$ (so also
$V_{12,i}=0$), relations (\ref{meas.error.mod}) and
(\ref{simp.reg.mod}) imply 
\bqa 
Y_{2i} &=& X_{2i} +\epsilon_{2i} \\
&=&\alpha_1 + \beta_1 X_{1i} + e_i + \epsilon_{2i} \\ 
&=& \alpha_1 + \beta_1 X_{1i} + e^*_i, 
\eqa 
where we have set $e_i^* = e_i+\epsilon_{2i}$. This is the typical setting 
for the application of WLS, provided that the variance of $e_i^*$ is 
independent of $Y_{2i}$. To do so, however, we need to estimate the 
variance of $e_i^*$. Note that, under the assumption made,
\bqa
V(e_i^*) = V(e_i) + V_{22,i}.
\eqa
Thus $V(e_i^*)$ is unknown because the intrinsic scatter $V(e_i)$ is 
unknown. We propose the following method for estimating $V(e_i)$.
 
\paragraph{Step 1.} Obtain $\hat \alpha_{OLS}, ~\hat\beta_{OLS}$ by a
direct application of OLS to the data $(Y_{2i},X_{1i})$.
 
\paragraph{Step 2.} Calculate the residuals
$$
R_i = Y_{2i}-\hat \alpha_{OLS} - \hat\beta_{OLS}X_{1i}.
$$
 
\paragraph{Step 3.} Obtain the estimator of $V(e_i)$ from
\bqan \label{est.V(e)}
\widehat {V(e_i)} = n^{-1}\sum_{i=1}^n(R_i-\bar R)^2
- n^{-1}\sum_{i=1}^nV_{22,i}.
\eqan
 
It can be shown that the estimator of $V(e_i)$ described in
(\ref{est.V(e)}) is consistent.  Next, set \bqan \label{est.V(e*)}
\widehat {V(e_i^*)} = \hat \sigma_i^{*2} = \widehat {V(e_i)} +
V_{22,i}, \eqan and let $A$ be the $n\times n$ matrix with
diagonal elements $\hat\sigma_i^{*2}$ and with all off-diagonal
elements equal to zero.  In terms of $A$, a general formula for the
WLS estimator is given in Arnold (1981). For the present
simple regression problem, this formula gives the following WLS
estimators for $\beta_1$,
\bqan \label{WLS.beta}
\hat\beta_{WLS}&=&\frac{\sum\hat\sigma_i^{*-2}
\sum\hat\sigma_i^{*-2}X_{1i}Y_{2i}
-\sum\hat\sigma_i^{*-2}X_{1i}\sum\hat\sigma_i^{*-2}Y_{2i}}
{\sum\hat\sigma_i^{*-2}
\sum\hat\sigma_i^{*-2}X_{1i}^2 -(\sum\hat\sigma_i^{*-2}X_{1i})^2} \\
\label{WLS.alpha}
\hat\alpha_{WLS} &=&
\frac{\sum\hat\sigma_i^{*-2}X_{1i}^2\sum\hat\sigma_i^{*-2}Y_{2i}
-\sum\hat\sigma_i^{*-2}X_{1i}\sum\hat\sigma_i^{*-2}X_{1i}Y_{2i}}{\sum
\hat\sigma_i^{*-2}\sum\hat\sigma_i^{*-2}X_{1i}^2
-(\sum\hat\sigma_i^{*-2}X_{1i})^2}.
\eqan
Variance estimates for the WLS estimators are
\bqan \label{var.WLS.beta}
\widehat V(\hat\beta_{WLS}) &=&
\frac{\sum\hat\sigma_i^{*-2}}{\sum\hat\sigma_i^{*-2}
\sum\hat\sigma_i^{*-2}X_{1i}^2
-(\sum\hat\sigma_i^{*-2}X_{1i})^2}\\
\label{var.WLS.alpha}
\widehat V(\hat\alpha_{WLS}) &=& \frac{\sum
\hat\sigma_i^{*-2}X_{1i}^2}{\sum
\hat\sigma_i^{*-2}\sum\hat\sigma_i^{*-2}X_{1i}^2
-(\sum\hat\sigma_i^{*-2}X_{1i})^2}.
\eqan
Note that these are conditional (given $X_{11},\ldots ,X_{1n}$) estimates
of the variance of the WLS estimators and, when the $\hat\sigma_i^*$ are
all equal, reduce to the usual variance estimates of the OLS estimator
(Draper \& Smith, 1981). 
 
\section{Other Estimators}

Text books in Statistics rarely mention a slope estimator other than
the OLS($X_2|X_1$). However, for some data sets in astronomy (and
other sciences) it is not clear which variable should be treated as
the independent and which as the dependent. Since OLS($X_1|X_2$) gives
a different slope, astronomers have invented the bisector slope.  This
corresponds to the line that bisects the OLS($X_2|X_1$) and
OLS($X_1|X_2$) lines. In addition, astronomers also use orthogonal
least squares (which finds the line that minimizes the squared
orthogonal distances) and some others. Astronomers are well aware
that, for any given data set, each of the above slopes will probably
be different. However, the concept that theses sample slopes estimate
different population slopes is more elusive. In other words,
the true OLS($X_2|X_1$) slope is different from the true
OLS($X_1|X_2$) slope as well as from all other true slopes. In still
different words, the OLS($X_2|X_1$) slope obtained from any given data
set is a biased estimator of the true OLS($X_1|X_2$) slope and all
other true slopes and, for large enough sample sizes, confidence
intervals around each of the sample slopes will not overlap. This
implies, of course, that there is no such thing as ``true slope'', but
there is a true OLS($X_1|X_2$) slope, a true bisector slope etc. This
point is raised in IFAB, but it is critical to explain again here:
Statistics cannot offer guidance as to which of the true slopes is
appropriate for any given situation. This decision is up to the
scientist.  Statistics offers variance formulas that indicate how
accurately each sample slope is estimating the corresponding true
slope. However, since these formulas are asymptotic (meaning valid for
large samples) simulation studies are recommended to examine the small
sample performance of each estimator.

In this section we present such slopes and intercepts for data
sets with measurement errors. Let $\hat\beta_2$, $\hat\beta_3$,
and $\hat\beta_4$ denote the OLS($X_1|X_2$), bisector, and orthogonal 
regression slope, respectively. The corresponding intercepts are
\bqan \label{interc.form}
\hat\alpha_\iota = \bar Y_2 - \hat\beta_\iota\bar Y_1 \ , \quad \iota = 2,3,4.
\eqan
The formula for the OLS($X_1|X_2$) slope is
\bqan \label{OLS(X$|$Y)}
\hat \beta_2 &=& \frac
{\sum_{i=1}^n(Y_{2i}-\bar{Y_2})^2 -\sum_{i=1}^n\Sigma_{22,i}}
{\sum_{i=1}^n(Y_{1i}-\bar{Y_1})(Y_{2i}-\bar{Y_2})
- \sum_{i=1}^n\Sigma_{12,i}} \ .
\eqan
The slopes $\hat\beta_3$, and $\hat\beta_4$ are given in terms of the 
slopes of $\hat\beta_1$, and $\hat\beta_2$ according to the formulas
given in Table 1 of IFAB. The asymptotic normality of all slope and 
intercept estimators follows by arguments similar to those in 
Appendix A of IFAB and the proof in the present Appendix. To estimate
the variance of each estimator, consider the notation. Let
\bqan \label{xi.for.beta_2}
\hat \xi_{2i} &=& \frac{(Y_{2i}-\bar Y_2)(Y_{2i}-\hat\beta_2 Y_{1i} -
\hat\alpha_2) +
\hat\beta_2V_{12,i}-V_{22,i}}{S_{Y_1,Y_2} - \bar V_{12}} \ , \\
\label{zeta.for.alpha_2}  
\hat \zeta_{2i} &=&Y_{2i} -\hat\beta_2 Y_{1i} -\bar Y_1\hat\xi_{2i}\ , \\
\label{xi.for.beta_3}
\hat\xi_{3i} &=& \frac{(1+\hat\beta_2^2)\hat\beta_3}{(\hat\beta_1+\hat\beta_2)
\sqrt{(1+\hat\beta_1^2)(1+\hat\beta_2^2)}}\hat\xi_{1i} +
\frac{(1+\hat\beta_1^2)\hat\beta_3}{(\hat\beta_1+\hat\beta_2)
\sqrt{(1+\hat\beta_1^2)(1+\hat\beta_2^2)}}\hat\xi_{2i}\ ,\\
\label{zeta.for.alpha_3}
\hat\zeta_{3i} &=&Y_{2i} -\hat\beta_3 Y_{1i} -E(Y_{1i})\hat\xi_{3i}\ , \\
\label{xi.for.beta_4}
\hat\xi_{4i} &=& \frac{\hat\beta_4}{\hat\beta_1^2
\sqrt{4+(\hat\beta_2 -\hat\beta_1^{-1})^2}}\hat\xi_{1i} +
\frac{\hat\beta_4}{
\sqrt{4+(\hat\beta_2 -\hat\beta_1^{-1})^2}}\hat\xi_{2i} \ ,\\
\label{zeta.for.alpha_4}
\hat\zeta_{4i} &=&Y_{2i} -\hat\beta_4 Y_{1i} -E(Y_{1i})\hat\xi_{4i} \ ,
\eqan
where $S_{Y_1,Y_2}=n^{-1}
\sum_{i=1}^n(Y_{1i}-\bar{Y_1})(Y_{2i}-\bar{Y_2})$, and 
$\bar Y_1$, $\bar Y_2$, $\bar V_{12}$ are the arithmetic averages of
$Y_{1i}$, $Y_{2i}$, $V_{22,i}$, respectively. Also, for 
$\iota = 2,3,4$, let $\hat\sigma_{\beta_\iota}^2$, 
$\hat\sigma_{\alpha_\iota}^2$, be the sample 
variances obtained from
$\hat\xi_{\iota i}, ~i=1,\ldots ,n$ and $\hat\zeta_{\iota i}, ~
i=1,\ldots ,n$, respectively. Arguing as in the Appendix,
the variances of $\hat\beta_\iota$ and $\hat\alpha_\iota$ are estimated by
\bqan \label{var.beta.alpha}
\widehat V(\hat\beta_\iota) = n^{-1}\hat\sigma_{\beta_\iota}^2 \ , \quad 
\widehat V(\hat\alpha_\iota) = n^{-1}\hat\sigma_{\alpha_\iota}^2 \ , 
\eqan
respectively, $\iota = 2,3,4$. In addition,  the covariance between
$\hat\beta_\iota$ and $\hat\alpha_\iota$ is estimated by
\bqan
\widehat{Cov}(\hat\beta_\iota,\hat\alpha_\iota) =
n^{-1}\hat\sigma_{\beta_\iota,\alpha_\iota} \ ,
\eqan
where $\hat\sigma_{\beta_\iota,\alpha_\iota}$ is the sample covariance
obtained from $(\hat\xi_{\iota i},\hat\zeta_{\iota i})$, $\iota = 2,3,4$.

We note in closing that the variance of
$\hat\beta_3$ is related to the variances of $\hat\beta_1$ and $\hat\beta_2$
through the formula given in Table 1 of IFAB. In particular,
\bqan \label{var2.bisec}
\hat V(\hat\beta_3) = \frac{\hat\beta_3^2}{(\hat\beta_1+\hat\beta_2)^2
(1+\hat\beta_1^2)(1+\hat\beta_2^2)}[(1+\hat\beta_2^2)^2
\widehat V(\hat\beta_1) &+& 2(1+\hat\beta_1^2)(1+\hat\beta_2^2)
\widehat{Cov}(\hat\beta_1,\hat\beta_2) \nonumber \\
&+&(1+\hat\beta_1^2)^2\widehat V(\hat\beta_2)].
\eqan
It can be shown analytically, that the value obtained from the formula 
in (\ref{var2.bisec})
will always be somewhat larger than that obtained from 
(\ref{var.beta.alpha}) with $\iota = 3$
but this difference will be negligible
for large sample sizes.  A similar remark 
applies for the variance of $\hat\beta_4$.

Some simulation studies examining the small sample performance of
these estimators and their estimated variances are presented in the 
next section in reference to a particular application.

\section{Example Applications To Real Data}

\subsection{Correlated errors and intrinsic scatter in $X_1$ and $X_2$}

\subsubsection{Color-luminosity relations}

Color-luminosity (CL) relations for galaxies have been characterized
by linear regressions of color (C) against absolute magnitude (M)
(Baum 1959). Often C and M both include the same band, so that their
errors are correlated. Most studies have also noted that the scatter
about the linear CL regression is larger than can be explained by
measurement error alone (e.g.\ Mobasher \etal 1986). Regressions for
this type of data, then, fall exactly in the domain of the models
developed in sections 2.1 and 3.

Almost without exception, studies of color-luminosity relations
have used OLS($X_2|X_1$), where M has been taken as the
independent variable. These regressions typically do not weight by
errors in $X_2$ (C), although several studies have included some
form of robust estimation via iterative rejection of outlying points
(Griersmith 1980, Bothun \etal 1985, Bower \etal 1992, and Bershady
1995). However, none of these studies have taken into account the
correlation in the errors of color and magnitude. Wyse (1982) avoided
this issue by fitting a linear regression directly to magnitudes in
two bands.

To assess the magnitude of the biases present in analyses using
incorrect statistical models, in Table 1 we compare a wide range of
linear regression models fit to CL relations for two subsets of data
from Bershady (1995). These subsets are defined as galaxies with
spectral types {\it bk}, and type {\it am} or {\it fm}, where the type
refers to the two stellar components dominating the galaxy's
broad-band colors. ``BCES'' models include bivariate, correlated
errors and intrinsic scatter (this paper). ``BES'' models include
bivariate errors and intrinsic scatter, but without the correlated
term $V_{12,i}$ (this paper). ``OLS'' models are those of IFAB,
which include only homoscedastic intrinsic scatter. Finally, ``WBE''
models, for the ($X_2|X_1$) case alone, include weighting by the
bivariate errors, but no error correlation or intrinsic scatter
(Bershady 1995). This method is formulated in terms of a minimizing
$\chi^2$, as defined by Press {\it et al.} (1988), and solved
numerically. \footnote{In Bershady (1995), this ``WBE'' method was
attributed to be similar to one described by Stetson (1989); it is,
however a substantially different statistical model.}  Bershady's
implementation of this method (used here) also includes iterative
rejection of highly deviant points (5 standard deviations in
error-normalized distance from the regression at each iteration). In
several regards, therefore, this model is significantly different than
any presented here. For each model the analytic estimates and standard
deviations for $\beta$ and $\alpha$ are listed on the first line, with
the results from 1000 simulations via bootstrap resampling on the
following line.

As might be expected from the shallow slope and substantial scatter in
the CL relation, the ($X_1|X_2$) regressions (and therefore the
bisectors) are steeper than the ($X_2|X_1$) regressions. More subtle
is the change (bias) with respect to models which include correlated
errors and intrinsic scatter: slopes become steeper for $(X_1|X_2)$
and bisector regressions and shallower for $(X_2|X_1)$ and orthogonal
regressions when correlated errors and intrinsic scatter are excluded
from the regression models. For each family of models, orthogonal and
($X_2|X_1$) regressions yield comparable slopes for these particular
data sets. The WBE($X_2|X_1$) model yields the shallowest slope.

What are the effects on possible scientific conclusions? If the CL
relation is to be understood physically (e.g. Arimoto \& Yoshii 1987),
then the ``BCES'' models should be used since they will give unbiased
results. However, the variances for ($X_2|X_1$) and orthogonal
regressions are comparable for all models, with the exception that the
boot-strap uncertainties are considerably smaller for {\it bk} type
galaxies for WBE($X_2|X_1$). (In contrast the variances for the
($X_1|X_2$) and bisector regressions become larger when intrinsic
scatter and measurement error are excluded from the statistical
model.)  As importantly, the regression slopes in these cases are the
same for the two samples for a given model.  One might therefore be
tempted to conclude that the OLS($X_2|X_1$) and WBE models adopted in
previous studies are satisfactory for comparisons of CL regression
slopes for different galaxy types (e.g. Mobasher \etal 1986 and
Bershady 1995, respectively) or at different redshifts (Stanford \etal
1995). While this appears to be the case for the particular data set
used here, in general OLS and WBE models yield biased results and
therefore their estimated variances are not {\it necessarily}
meaningful quantities since they do not include the effects of the
{\it unknown} bias. Hence BCES models should be used even for slope
comparisons between samples.

When using CL relations to estimate distance moduli (e.g. Sandage
1972), zeropoint differences between samples may be better estimated
using cross-correlation techniques (e.g.\ Dressler 1984), as done by
Bower \etal (1992). 

Within the ``BCES'' family of models, which regression is best to use
for CL-relation studies?  To understand the bias and accuracy (due to
small numbers) of each of these regressions, we conducted two
simulation studies with artificially generated data sets designed to
closely match the above observed color-luminosity distributions.  One
set of simulations, $(X_{1i},X_{2i}),$ $i=1, \ldots ,n$, was generated
according to the model in (\ref{simp.reg.mod}) with $\alpha_1 = 2.5$,
and $\beta_1 = 0.07$. The range of the $X_1$-values was (-28, -18) and
the intrinsic scatter was generated according to a normal distribution
with zero mean and standard deviation 0.55. Normal measurement errors
were added to the $(X_1,X_2)$-values in order to simulate the observed
data $(Y_{1i},Y_{2i}),$ $i=1,\ldots ,n$. The range for the variances
of the measurement errors was (0.18, 0.45), with the covariance fixed
at 0.15.  A second set of simulations used $\beta_1$=0.12 0.12, a
measurement error on $X_1$ with variance ranging from (0.03,0.3), a
range of variance of the measurement error on $X_2$ of (0.06,0.6), and
a range of the covariance of the two measurement errors of (0.03,0.3);
all other parameters were the same as before. For both simulation
sets, the randomly generated data $(Y_{1i},Y_{2i}),V_i$ were fed
into the BCES routine and the entire process was repeated 1000 times.
The recorded outcome was the average (over the 1000 simulation runs)
of the estimated coefficients, the sample variance of the the 1000
estimated coefficients, and the average value of the variance formulas
for each of the estimated coefficients. Samples of $n=50$, 150 and 500
were generated to understand the effects of small sample sizes on the
estimated coefficients and variances.

For these simulation studies, $\hat\beta_1$ (BCES($X_2|X_1$))
performed best in all respects: Even with $n=50$ the bias
(small-sample bias) was small and the sample variance over the 1000
simulations closely matched the average variance computed from the
formula. The variance of $\hat\beta_1$ was the smallest of all the
estimators (a factor of 4 better than the next smallest variance).
There was no noticeable change in the performance of BCES($X_2|X_1$)
for the different sample sizes. The performance of the other
estimators did change with the sample size. When the true $\beta_1$
slope was 0.07, BCES($X_1|X_2$) and BCES-orthogonal regressions had
considerable biases in their slopes $\hat\beta_2$ and $\hat\beta_4$
for $n=50$. When $\beta_1$ was set to 0.12, BCES-orthogonal regression
slope, $\hat\beta_4$, performed better than BCES($X_1|X_2$) and
BCES-bisector regressions slopes $\hat\beta_2$ and $\hat\beta_3$ in
terms of both bias and variance for $n=150$ and 500.  On the basis of
these simulation results we recommend the use of BCES($X_2|X_1$) for
color-luminosity data sets similar to those presented here. An
important caveat is that different model specifications might result
in different performance of the estimators. This should be checked for
each specific study.

\subsubsection{The Tully-Fisher relation}

Another example of data with correlated errors is the relation between
spectral line-width (internal velocity) and luminosity of spiral
galaxies (Tully \& Fisher 1977). Here too there exists a dispersion
about the linear regression in addition to measurement error (e.g.\
Pierce \& Tully 1992). The error correlation occurs because both the
velocity (corrected for projection) and absolute magnitude (corrected
for dust extinction) depend on the inferred inclination. There can be
non-negligible uncertainties in the inclination measurement,
particularly for galaxies that are not spatially well-resolved.  The
effect of the correlated magnitude and velocity errors will be to
increase the observed scatter and hence make the observed relation too
flat {\it if} unaccounted for. In all cases, linear regressions should
be computed for the Tully-Fisher relation using the BCES model in
preference over other existing models.

We note however, that the BCES model is not immune to biases that
arise from data truncation (i.e. a luminosity-limited sample, or a
magnitude-limited sample in the case of a cluster). This has been
explored in detail by Willick (1994, and references therein) for the
case of intrinsic scatter but uncorrelated errors. Another limitation
of the BCES model is that it does not allow for changes in the scatter
along the regression. The sample of Mathewson \etal (1992) suggests
that the scatter in the current Tully-Fisher relation (as defined by
21-cm integrated line-widths) increases at lower velocities or
luminosities. Future work should consider elaborations of the BCES
statistical model to include variable intrinsic scatter, as well as
estimation of this scatter.

\newpage

\subsection{Errors in $X_2$ only and intrinsic scatter: The Tolman Test}

A fundamental test in observational cosmology is verification that
redshift is caused by a secular change in the metric (Tolman 1930),
namely universal expansion. If true, then surface-brightness in a
fixed band-pass scales as the kinematic factor $(1+z)^{-3}$,
independent of other cosmological parameters (although for
astrophysical sources such as galaxies, the dimming is modified by the
$K$-correction). There have been several recent attempts to perform
the Tolman test (Sandage \& Perelmuter 1990, Kjaergaard, Jorgensen \&
Moles 1993, and Pahre, Djorgovski \& de Carvalho 1996).  These studies
note that the {\it rest-frame} surface-brightnesses of galaxies are
not all the same, but depend on a number of variables, including
luminosity or size. However, even after taking this into account,
galaxy samples are likely to still have some intrinsic dispersion in
surface-brightness at a given redshift. However, in terms of
measurement errors, redshifts can typically be measured with high
precision compared to the apparent magnitudes and sizes needed to
derive surface-brightnesses.  Hence, the linear regression for
surface-brightness vs. log(1+z) for such a data set is well
approximated by the WLS model presented here.\footnote{Any linear
correlation as a function of redshift is likely to fall in this
category for astrophysical sources.}  In fact, since the WLS model
includes (and estimates) intrinsic scatter, it is not necessary to
``reduce'' the data to some fiducial surface-brightness on the basis
of luminosity or size. This has the distinct advantage of allowing the
test to be extended to galaxies for which tight ``Fundamental Plane''
relations have not yet been formulated.

We have tested the WLS method with two simulation studies. The first
simulated data sets were generated as described in the simulations
reported in subsection 4.2.1, but without adding the error in the
$X_1$ variable. The small-sample bias of the WLS estimator was
comparable to that of BCES($X_2|X_1$), but the variance of the WLS
estimator was an order of magnitude smaller than that of
BCES($X_2|X_1$)! The same results were found for the second simulated
data set with parameters designed to mimic the Tolman test in the $K$
band, assuming surface-brightnesses are measured in large, metric
apertures to redshifts of $\sim$0.4, and $K$-corrected
surface-brightnesses are plotted versus 2.5 log(1+z): $\alpha_1$ = 16,
$\beta_1$ = 4, $X_1$ in the range (0, 0.4), intrinsic scatter given by
a normal distribution with zero mean and standard deviation of 0.3,
and normal measurement errors on $X_2$ with variances in the range
(0.03, 0.3). However, the formula for the confidence interval on
$\hat\beta_1$ for this second study gave conservative results
(i.e. wider confidence intervals) even for sample sizes of 500. Only
for sample sizes of 900 did the formula for the confidence interval
capture the true variability of the WLS estimator. This may be due to
the narrow range of the $X_1$-values (the confidence interval formula
performed well for much smaller numbers for the first simulation set).

On the basis of these simulation results we recommend the use of the
WLS estimator whenever the $X_1$ variable is observed without
measurement error and the magnitudes of the measurement errors can be
assumed independent from the observations. This would be case for the
Tolman test, for example, when flux measurements are
background-limited. As an additional bonus, the WLS provides an
estimate of the intrinsic scatter. However, for a narrow range of
$X_1$-values, we recommend the use of bootstrap confidence intervals
even for relatively large sample sizes.

\section{Discussion}

To our knowledge, the methods presented here are the only algorithms
that apply to data with both measurement errors and intrinsic
scatter. When is it necessary to use one of the above methods over the
techniques discussed in IFAB or FB? There are two basic criteria for
selecting a statistical model to use for studying correlations in
data, bias and uncertainty. Their relative importance depends somewhat
on the specific scientific objective, however the conclusion is the
same: When in doubt, the BCES and WLS models should be used. However,
the WLS model should be used {\it only} in the approximation where the
$X_1$ variable is measured without error.

If the purpose is to test a theory which predicts correlation slopes
and/or zeropoints for some set of observables, then bias is the
principal criterion. The statistical model which best approximates the
real data is expected to give the least-biased regression, and so the
choice becomes an issue of approximation. Because astronomy largely
consists of passive observations and not active experiments, there is
rarely an `explanatory' variable free of measurement error. Moreover,
correlations between variables for astronomical systems almost always
have intrinsic scatter, which is simply a reflection of these systems'
complex, multi-variate dependencies. The `Fundamental Plane' for
elliptical galaxies is one good astronomical example of this
complexity (cf.\ Santiago \& Djorgovski 1993). For cases where the
intrinsic scatter may be {\it much larger} than measurement error, or
vice-versa, the methods in IFAB or those outlined in FB, respectively,
may provide acceptable approximations. However, at this time we cannot
quantify ``much larger''. The methods presented here are valid in
general and, since they reduce to the methods considered in IFAB in
the case of no measurement errors, {\it we recommend that the present
methods be used in all cases.}

There are some situations where differential measurements are designed
simply to detect differences in slope between samples. Examples of
this were described for the CL relation.  Here, the most accurate
regression estimate may be desired, and should be assessed via
simulations of artificial data sets, as we have illustrated for the
BCES family of models. However, if the statistical model is incorrect,
then the estimated variance does not necessarily include effects of
bias, which may differ from sample to sample.  While bootstrap
estimates of the variance may be 'unbiased,' the same is not
necessarily true of the slope. To put it another way, if the null
hypothesis is that two samples are the same, and this is to be
confirmed by comparing regression properties, any statistical model
may yield results consistent with the hypothesis. However, if the
statistical model is incomplete or unrepresentative of the data, the
comparison is only consistent and cannot validate the
hypothesis. Again, BCES models are the most general and should provide
the least-biased estimates of regression slopes and variances.

Within a family of regressions models (e.g.\ BCES or OLS), the choice
of particular regression (($X_2|X_1$), ($X_1|X_2$), etc.) is only an
issue of accuracy, and {\it not} bias.  As has been emphasized in IFAB
and again here, {\it the different regression methods give different
slopes even at the population level.} All slopes are related to the
second moments of the bivariate distribution of the data. Again, the
most accurate regression should be assessed via simulations.

In the case where the $X_1$ variable is measured without error, our
simulations for two different artificial data sets revealed that the
WLS estimator has smaller variance than BCES($X_2|X_1$). However WLS
is consistent only when the error magnitude is independent from the
observation. While the BCES estimators are consistent under general
conditions, the simulations suggest they can be improved under the
additional assumption that the measurement errors on $X_1$, $X_2$ are
independent from the observations. Weighted versions of the BCES
estimators under this additional assumption will be the subject of a
forthcoming paper.
 
The present procedures resulted from an interdisciplinary
collaboration of astrophysicists and mathematical statisticians via
the newly founded {\it Statistical Consulting Center for Astronomy}
(SCCA). Further information about SCCA can be obtained through the
World Wide Web (http://www.stat.psu.edu/scca/homepage.html), or by
contacting SCCA@stat.psu.edu. A FORTRAN package which includes the
algorithms in this paper and IFAB, including bootstrap resampling
error analysis, is available via anonymous ftp (contact
mab@astro.psu.edu).

The work of MGA was supported in part by NSF grant DMS-9208066. MAB
acknowledges support from NASA through grant HF-1028.02-92, from the
Space Telescope Science Institute, which is operated by the
Association of Universities for Research in Astronomy, Incorporated,
under contract NAS5-26555.

\appendix
\section{Proofs}

{\bf Proof of Relations (\ref{mean.rel}), (\ref{var.rel}) and 
(\ref{cov.rel}).}  Relation (\ref{mean.rel}) is obvious from 
relation (\ref{meas.error.mod}) and the fact that, conditionally on 
$V_{kk,i}$, the errors $\epsilon_{ki}, ~k=1,2$ have zero mean. 
To show (\ref{var.rel}), note that
\bqa
E(Y_{ki}^2)&=& E[E(Y_{ki}^2|V_{kk,i})] \\
&=& E[E((Y_{ki}-X_{ki})^2 +X_{ki}^2 +2X_{ki}(Y_{ki}-X_{ki})|V_{kk,i})] \\
&=& E[E(\epsilon_{ki}^2 +X_{ki}^2 +2X_{ki}\epsilon_{ki}|V_{kk,i})] \\
&=& E(V_{kk,i}) + E(X_{ki}^2).
\eqa
Since the variance of any random variable $Z$ is $V(Z)=E(Z^2)-[E(Z)]^2$, 
(\ref{var.rel}) follows from the above relation and (\ref{mean.rel}).
Similarly, the proof of (\ref{cov.rel}) follows from
\bqa
E(Y_{1i}Y_{2i}) &=& E[E(Y_{1i}Y_{2i}|\vV_{ki})] \\
  &=& E[E(\epsilon_{1i}\epsilon_{2i} + X_{1i}X_{2i} +X_{1i}\epsilon_{2i}
+ X_{2i}\epsilon_{1i}|\vV_{ki})] \\
 &=& E(V_{12,i}) + E(X_{1i}X_{2i}),
\eqa
the fact that the covariance of any two random variables $Z_1,Z_2$, is
$Cov(Z_1,Z_2)=E(Z_1Z_2)-E(Z_1)E(Z_2)$ and from (\ref{mean.rel}).

\bigskip

\noindent{\bf Proof of Asymptotic Normality of $\hat\beta_1$.} Write 
$S_{Y_1,Y_2}=n^{-1}
\sum_{i=1}^n(Y_{1i}-\bar{Y_1})(Y_{2i}-\bar{Y_2})$, and $S_{Y_1}^2=
n^{-1}\sum_{i=1}^n(Y_{1i}-\bar{Y_1})^2$. We will need the following 
relations.
\bqan \label{scov.exp}
\sqrt{n}(S_{Y_1,Y_2}-C(Y_{1},Y_{2}))& =& n^{-1/2}
\sum_{i=1}^n(Y_{1i}Y_{2i}-E(Y_{1}Y_{2})) - E(Y_{1}) 
n^{1/2}(\bar{Y_2}-E(Y_{2}))\\
&-&E(Y_{2})n^{1/2}(\bar{Y_1}-E(Y_{1})) + o_p(1),\nonumber \\ \label{svar.exp}
\sqrt{n}(S_{Y_1}^2 - V(Y_{1}))&=& n^{-1/2}\sum_{i=1}^n(Y_{1i}^2- E(Y_{1}^2))
- 2E(Y_{1})n^{1/2}(\bar{Y_1}-E(Y_{1})) \\
& +&o_p(1), \nonumber
\eqan
where $o_p(1)$ denotes a quantity that converges to zero in probability as
$n\rightarrow \infty.$ Write
\bqan \label{beta1.rep}
\sqrt{n}(\hat\beta_1-\beta_1)&=&\sqrt{n}\left[\frac{S_{Y_1,Y_2}-
\bar{V}_{12}}{S_{Y_1}^2 -\bar{V}_{11}} -\frac{C(Y_{1i},Y_{2i}) - 
E(V_{12,i})}{V(Y_{1i}) - E(V_{11,i})}\right]\\
&=&\sqrt{n}\left[\frac{S_{Y_1,Y_2}-C(Y_{1i},Y_{2i})- (\bar{V}_{12} -
E(V_{12,i}))}{V(Y_{1i}) - E(V_{11,i})} \right.\nonumber \\
&-&\left. [C(Y_{1i},Y_{2i}) - 
E(V_{12,i})]\frac{S_{Y_1}^2 -V(Y_{1i}) -(\bar{V}_{11}
-E(V_{11,i})}{[V(Y_{1i}) - E(V_{11,i})]^2}\right] + o_p(1)
\nonumber
\eqan

Using (\ref{scov.exp}) and (\ref{svar.exp}), it can be seen after some
algebra that (\ref{beta1.rep}) can be written as
\bqa
\sqrt{n}(\hat\beta_1-\beta_1)&=&\sqrt{n}(\bar{\xi_1}-E(\xi_{1i})) + o_p(1),
\eqa
and an application of the Central Limit Theorem completes the proof of the 
asymptotic normality.
That (\ref{sigma.beta.hat}) and (\ref{sigma.alpha.hat}) provide 
consistent estimators of $\sigma^2_{\beta_1}$ and $\sigma^2_{\alpha_1}$ is
straightforward.

\bigskip

\noindent {\bf Acknowledgment.} Thanks are due to J. Willick for
many constructive comments that greatly improved the presentation of 
the paper.

\clearpage

\begin{figure}
\plotfiddle{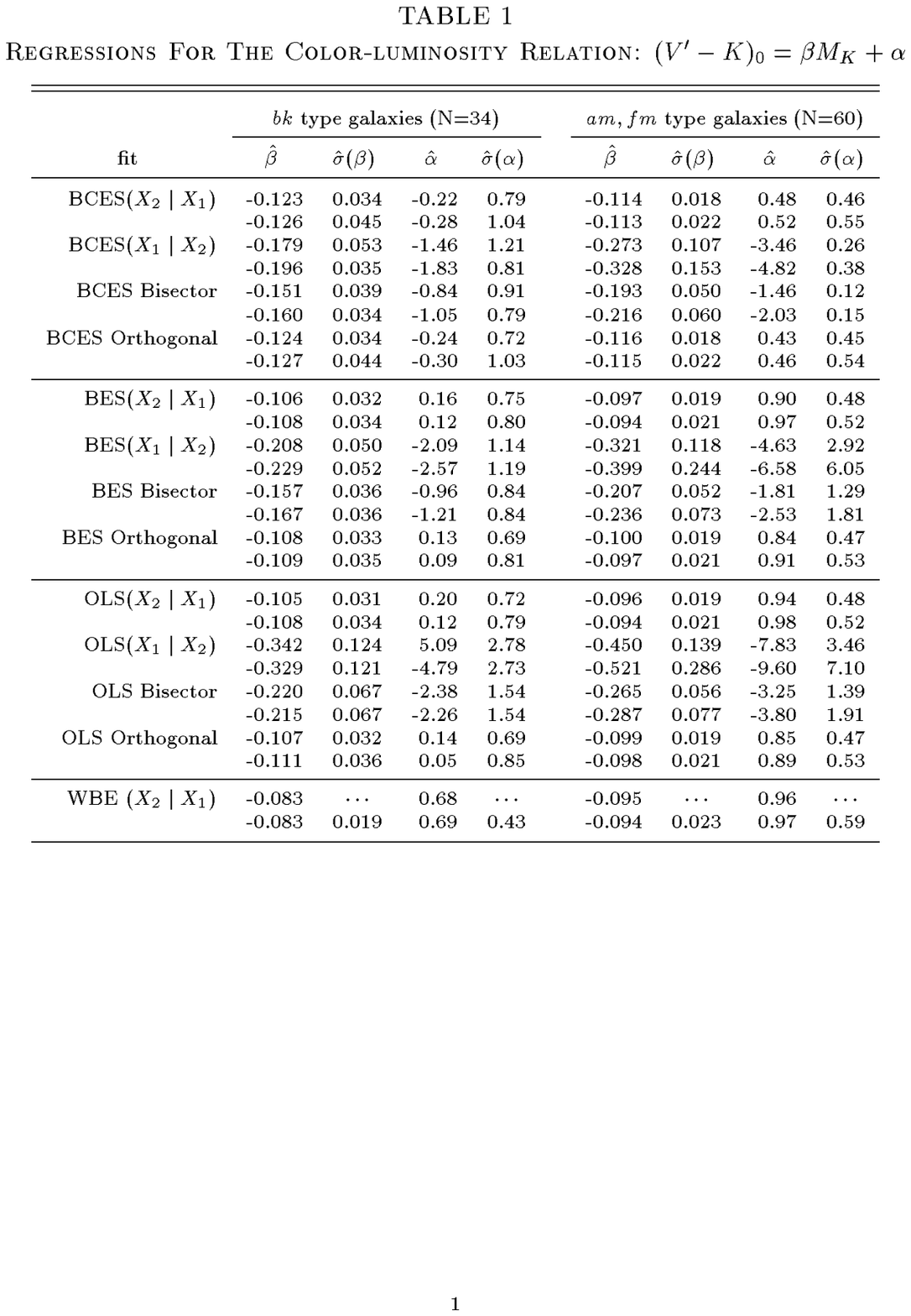}{8in}{0}{100}{100}{-310}{-100}
\end{figure}

\end{document}